\documentclass[PoP,twocolumn,showpacs,%
preprintnumbers,amssymb,prb]{revtex4-1}
\usepackage{graphicx}
\usepackage[colorlinks=true,linkcolor=blue,%
linkcolor=blue,pagecolor=blue,citecolor=blue,%
urlcolor=blue,anchorcolor=black,%
bookmarksnumbered=true,%
bookmarksopen=true]{hyperref}
\usepackage{dcolumn}
\usepackage{bm}
\renewcommand{\frac}[2]{\displaystyle{#1 \over #2}}
\begin{document}
\title{A SCALING LAW FOR THE DUST 
CLOUD IN RF DISCHARGE UNDER 
MICROGRAVITY CONDITIONS}
\author{D.~I.~Zhukhovitskii} \email{dmr@ihed.ras.ru}
\affiliation{Joint Institute of High Temperatures, Russian 
Academy of Sciences, Izhorskaya 13, Bd.~2, 125412 
Moscow, Russia}
\author{V.~I.~Molotkov}
\affiliation{Joint Institute of High Temperatures, Russian 
Academy of Sciences, Izhorskaya 13, Bd.~2, 125412 
Moscow, Russia}
\author{V.~E.~Fortov}
\affiliation{Joint Institute of High Temperatures, Russian 
Academy of Sciences, Izhorskaya 13, Bd.~2, 125412 
Moscow, Russia}
\date{\today}
\begin{abstract}
We employ the approximation of overlapped 
scattering potentials of charged dust particles 
exposed to streaming ions to deduce the 
``equation of state" for a stationary dust cloud 
in the radio frequency (RF) discharge apart 
from the void--dust boundary. The obtained 
equation defines the potential of a dust particle 
as a function of the ion number density, the 
mass of a carrier gas atom, and the electron 
temperature. A scaling law that relates the 
particle number density to the particle radius 
and electron temperature in different systems is 
formulated. Based on the proposed approach 
the radius of a cavity around a large particle in 
the bulk of a cloud is estimated. The results of 
calculation are in a reasonable agreement with 
the experimental data available in literature.
\end{abstract}
\maketitle
\section{\label{s1} INTRODUCTION}

Dusty or complex plasma is a low-temperature plasma, 
which includes dust particles with sizes ranging from $1$
 to $10^3 \;\mu {\mbox{m}}$. Due to the higher electron 
mobility, particles acquire a considerable electric charge. 
Thus, a strongly coupled Coulomb system is 
formed.\cite{1,2,3,4,5,6,8,9} In such a plasma, various 
collective phenomena at the level of individual particles can 
be observed. Complex plasmas are usually studied in gas 
discharges at low pressures, e.g., in the radio frequency 
(RF) discharges. Under microgravity conditions, a large 
homogeneous bulk of the complex plasma can be observed. 
The microgravity conditions are realized either in parabolic 
flights\cite{10,11,12,13,14} or onboard the International 
Space Station (ISS).\cite{10,15,16,17,18,19}

In most studies, attention is focused on individual particles 
injected in plasma, which is a model of the rarefied dust 
cloud. Elaborate theories model the charge of a particle and 
its screening by the plasma, elementary processes occurring 
on the particle surface, interaction between the particles and 
streams of ions and neutrals associated with the momentum 
transfer, interaction of particles at large distances 
etc.\cite{1} At the same time in the dust clouds observed 
experimentally, dust particles are situated at the distances, 
which are insufficiently large to neglect collective 
phenomena that control the particle charging and stability 
of a dust cloud. The strong interaction between dust 
particles and their interaction with the ions and electrons 
result in such a collective phenomenon as the void 
formation. Formation and stability of the voids were 
investigated in 
Refs.~\onlinecite{26,29,30,31,32,33,34,35,36,37,38}. 
These studies, however, were aimed at the void--dust 
boundary, and little attention was paid to the parameters of 
the bulk of a dust cloud apart from the boundary.

Another manifestation of the collective phenomena in a 
dust cloud is a spatial distribution of the particles. It was 
noticed long ago that a dust cloud with an arbitrary particle 
number density could not be created for the given particle 
radius. The number of particles injected into the RF 
discharge plasma influences solely the volume occupied by 
the dust cloud rather than the particle number density. In 
different experiments, the latter seems to scale with the 
particle radius and electron temperature in some way. This 
phenomenon is not understood so far.

Some experiments are carried out on an inhomogeneous 
system consisting of the particles with different diameters. 
The simplest example of such a system is a large particle 
surrounded by a dense cloud of smaller particles. Usually, 
this particle called the projectile moves through the cloud 
with a supersonic or subsonic velocity. Such projectiles are 
generated using controlled mechanisms of 
acceleration\cite{11,20}, or they can appear 
sporadically.\cite{19,21} A strongly coupled Coulomb 
system like dust particles in the gas discharge plasma can 
be represented as a system of the Wigner--Seitz cells with a 
particle in the center of each cell. According to a natural 
assumption of the cell quasineutrality and to the 
proportionality between the particle radius and its charge, 
the particle diameter must be proportional to the cube of 
cell radius. Instead, it was observed experimentally that the 
radius of a cell around a projectile is always noticeably 
larger than it could be expected from the foregoing 
considerations. This fact was also not understood as yet.

We propose a model of a dust cloud based on the allowance 
for a collective interaction between a dense cloud of the 
dust particles and the streaming ions. As it was noted in an 
early study,\cite{26} stability of a particle in a dust cloud is 
provided by the balance between the ion drag force and the 
electric force generated due to the ambipolar diffusion. We 
point to the fact that a dust cloud realized in the experiment 
is so dense that the characteristic impact parameter of the 
momentum transfer from the ions to a particle is {\it 
larger\/} than the interparticle distance. In other terms, the 
amount of the moment transfer is restricted by the overlap 
of scattering potentials of neighboring particles. As a 
consequence, the drag force turns out to be dependent on 
the particle number density. With due regard for this fact 
we construct the ``equation of state" for the dust cloud 
based on the equation of force balance, the equation 
defining the particle charge, and the overall plasma charge 
balance (quasineutrality). This equation of state is written in 
dimensionless quantities reduced to their ``critical" values 
similar to those appearing in the van der Waals equation. 
Based on this equation, one can calculate the particle 
number density as a function of the ion number density, the 
electron temperature, and the mass of a carrier gas atom. 
The obtained equation makes it possible to formulate a {\it 
scaling law\/} for a homogeneous dust cloud, which states 
that for the same carrier gas, the ratio of the squared 
interparticle distance to the product of the particle radius 
and the electron temperature must be constant in different 
systems. If the ion number density is sufficiently large, the 
equation of state has no solutions. Apparently, this is 
indicative of the existence of the void--dust boundary.

The equation of state allows one to treat the simplest 
inhomogeneous system, namely, a projectile in the dust 
cloud. We find the radius of a cavity around the projectile 
from a balance between the changes of the work against the 
static pressure of dust particles required to create a cavity 
and the energy of the electric field induced by the charge of 
a projectile. The static pressure is calculated from the 
equation of state of dust particles. The resulting cavity 
radius proves to be proportional to the square root of the 
projectile radius rather than to the power of 1/3.

The paper is organized as follows. In Sec.~\ref{s2}, the 
equation of state is deduced for a homogeneous dust cloud. 
The radius of a cavity around a large particle in such a 
cloud is estimated in Sec.~\ref{s3}. Calculation results are 
compared with available experimental data in 
Sec.~\ref{s4}, and the results of this study are summarized 
in Sec.~\ref{s5}.

\section{\label{s2} A STATIONARY HOMOGENEOUS 
DUST CLOUD}

Consider a stationary dust cloud in the RF discharge and 
formulate the condition of a particle mechanical 
equilibrium. The stability of such a state will not be tested. 
Consequently, the condition formulated in what follows 
may refer both to a stable and to an unstable branch of 
plasma parameters. Under microgravity conditions, a 
quiescent dust particle is subject to only two forces, 
namely, the force acting from the electric field $F_e = eZ_d 
E$
 and the ion drag force $F_{id} = \sigma _{\mathrm{eff}} 
n_i m_i v_{T_i } u_i $. Here, $Z_d = (a_d T_e /e^2 )\Phi 
_d$ is the dust particle charge in units of the elementary 
charge $e$, $a_d$ is the dust particle radius, $T_e$ is the 
electron temperature, for which we use the energy units 
($k_B = 1$), and $\Phi _d = e\varphi _d /T_e$ is the 
dimensionless potential of a dust particle, $\varphi _d$ is its 
potential; $E$
 is the electric field strength, $\sigma _{\mathrm{eff}}$ is 
the effective cross section of a collision between the ion 
and the particle, $n_i$ is the number density of ions inside 
the dust cloud, $m_i$ is the ion mass, $v_{T_i } = (T_i 
/m_i )^{1/2}$ is the ion thermal velocity, and $u_i$ is the 
ion drift velocity, which is implied to be much smaller than 
$v_{T_i } $. In a real system, $u_i \sim v_{T_i }$ but the 
expression for the drag force is too complicated in this case 
for our purpose to make analytical estimates. Since this 
general expression shows a regular behavior in the region 
$u_i \sim v_{T_i }$\cite{26}, it will not change our 
estimations qualitatively if we use the approximation $u_i 
\ll v_{T_i } $.

The ion drag force arises due to scattering of the streaming 
ions on dust particles. For an isolated dust particle, the 
Coulomb cross section of the momentum transfer from an 
ion to a particle is proportional to $Z_d^2$ and can be 
estimated in the order of magnitude as $(a_d \tau \Phi _d 
)^2 $, where $\tau = T_e /T_i $.\cite{37} For typical 
parameters of a dust cloud in the RF discharge produced on 
PK-3 Plus setup,\cite{19} $\tau \Phi _d \sim 300$, so that 
$\sigma _{\mathrm{eff}}^{1/2}$ exceeds noticeably the 
average distance between the dust particles characterized by 
the radius of the Wigner--Seitz cell for dust particles $r_d = 
(3/4\pi n_d )^{1/3} $. Under these conditions, the potentials 
acting on an ion from the neighboring particles overlap. 
This situation is typical, e.g., for the dense plasma of alkali 
metal vapors,\cite{39} where the electron mobility is 
defined by the overlapped potentials of individual scattering 
atoms. The overlapping reduces $\sigma _{\mathrm{eff}}$ 
dramatically, and it can be estimated as $Cr_d^2 $, where 
$C$
 is some coefficient. If $r_d \lesssim \lambda _{D_i } $, 
where $\lambda _{D_i } = (T_i /4\pi n_i e^2 )^{1/2}$ is 
the ion Debye length, the potential of a particle can be 
approximated by the unscreened Coulomb potential in the 
most part of a cell except for the vicinity of its boundary. 
The quasineutrality of the Wigner--Seitz cell leads to an 
effective cutoff of the Coulomb potential inside the cell at 
the distance $ \simeq 0.45r_d$ (see, e.g., 
Ref.~\onlinecite{50}). We will assume that the ion--particle 
scattering is equivalent to the collisions of the ions against a 
hard sphere with the radius $0.45r_d $. Then the 
corresponding drag force can be written as $F_{id} = (\pi 
/2)r_d^2 n_i \lambda _{in} eE$, where it has been taken 
into account that the ion drift velocity in the electric field 
$E$
 is $u_i = (\lambda _{in} /m_i v_{T_i } )eE$, where 
$\lambda _{in}$ is the ion mean free path with respect to 
the collisions against neutrals.

At the equilibrium, $F_e = F_{id} $, and we obtain the 
force balance equation
\begin{equation}
\frac{\pi }{2}r_d^2 n_i \lambda _{in} = \frac{{a_d T_e 
}}{{e^2 }}\Phi _d . \label{e1}
\end{equation}
The particle potential will be defined using the orbital 
motion limited (OML) approximation, which proves to be 
useful in most applications.\cite{40,41} The quantity $\Phi 
_d$ is a solution of the equation\cite{1}
\begin{equation}
e^{\Phi _d } (1 + \tau \Phi _d ) = \frac{{n_e }}{{n_i 
}}(\tau \mu )^{1/2} , \label{e2}
\end{equation}
where $\mu = m_i /m_e $. Note that $r_d$ is typically 
greater or much greater than $a_d (\tau \Phi _d )^{1/2} $, 
which means that the OML approximation is still valid, i.e., 
the ion flux on a particle is almost the same as for an 
isolated particle, unless the scattering potentials overlap. 
This is due to the fact that large impact parameters $ \sim 
r_d$ contribute appreciably to the total Coulomb moment 
transfer cross section.

Equations (\ref{e1}) and (\ref{e2}) are completed by the 
local quasineutrality condition
\begin{equation}
n_i = \frac{{aT_e }}{{e^2 }}\Phi _d n_d + n_e , \label{e3}
\end{equation}
which is accurate to $\lambda _{D_e } /L$, where 
$\lambda _{D_e } = (T_e /4\pi n_e e^2 )^{1/2}$ is the 
electron Debye length and $L$
 is the system length scale. We introduce the dimensionless 
quantities $n_i^* = (e^2 \lambda _{in}^3 /a_d T_e )n_i $, 
$n_e^* = (e^2 \lambda _{in}^3 /a_d T_e )n_e $, and 
$n_d^* = (4\pi /3)\lambda _{in}^3 n_d$ to reduce 
Eqs.~(\ref{e1})--(\ref{e3}) to a transcendental equation 
with respect to $\Phi _d$ provided that $\tau $, $\mu $, and 
$n_i^*$ are treated as parameters
\begin{equation}
\frac{{e^{\Phi _d } (1 + \tau \Phi _d )}}{{\tau ^{1/2} \mu 
^{1/2} }} + \frac{3}{8}\left( {\frac{{\pi n_i^* }}{{2\Phi 
_d }}} \right)^{1/2} = 1. \label{e4}
\end{equation}
Relation (\ref{e4}) is a scaled ``equation of state" for the 
dust cloud in the RF discharge. It defines the dust particle 
number density
\begin{equation}
n_d^* = \left( {\frac{{\pi n_i^* }}{{2\Phi _d }}} 
\right)^{3/2} \label{e5}
\end{equation}
and the electron number density
\begin{equation}
n_e^* = \frac{{e^{\Phi _d } (1 + \tau \Phi _d )}}{{(\tau 
\mu )^{1/2} }}n_i^* . \label{e6}
\end{equation}
From Eqs.~(\ref{e4})--(\ref{e6}), it is seen that it is 
impossible to create a dust cloud with an arbitrary particle 
number density by a simple addition of the particles to the 
cloud; instead, $n_d$ is defined by the system parameters.

Approximate solutions of Eq.~(\ref{e4}) can be obtained 
for the limiting cases $H \ll 1$
 and $H \gg 1$, where $H = \left| {Z_d } \right|n_d /n_e$ is 
the Havnes parameter.\cite{1} In the first case, $n_e \simeq 
n_i $, and $\Phi _d = \Phi _0 $, where $\Phi _0$ is the 
solution of the equation
\begin{equation}
e^{\Phi _0 } (1 + \tau \Phi _0 ) = (\tau \mu )^{1/2} , 
\label{e006}
\end{equation}
and we have
\begin{equation}
n_d^* = \left( {\frac{{\pi n_i^* }}{{2\Phi _0 }}} 
\right)^{3/2} . \label{e005}
\end{equation}
Here, an approximate solution of Eq.~(\ref{e006}) is $\Phi 
_0 \simeq (1/2)\ln (\mu /\tau ) - 1$.

If the dust particles number density (\ref{e005}) is 
sufficiently low so that the condition $H \ll 1$
 is satisfied, one can approximate the spatial ion number 
density distribution by that in the RF discharge without 
particles. For the sake of simplicity, we will assume that 
inside a dust cloud, the overall recombination rate enhanced 
due to recombination on the particle surface is nearly 
compensated by the ionization rate inhibited by the reduced 
electron number density. Then the continuity equation for 
the ions reads
\begin{equation}
\nabla \cdot (n_i {\bf{u}}_i ) = 0. \label{e050}
\end{equation}
The ion flux velocity can be estimated as ${\bf{u}}_i = 
(e\lambda _{in} /m_i v_{T_i } ){\bf{E}}$, where 
${\bf{E}} = (T_e /e)\nabla \ln n_e \simeq (T_e /e)\nabla \ln 
n_i$ is the electric field strength, and we arrive at
\begin{equation}
\nabla ^2 n_i = 0. \label{e060}
\end{equation}
Assuming the spherical symmetry of the discharge we 
obtain
\begin{equation}
n_i (r) = \frac{{n_{i0} r_0 }}{r},\quad n_{i0} = n_i (r_0 ), 
\label{e070}
\end{equation}
where $r$
 is the radial coordinate.

Consider the opposite case $H \gg 1$. Here, the 
quasineutrality equation (\ref{e3}) is reduced to
\begin{equation}
n_i \simeq \left| {Z_d } \right|n_d = \frac{3}{{4\pi 
}}\frac{{a_d T_e \Phi _d }}{{e^2 r_d^3 }}. \label{e7}
\end{equation}
It follows from (\ref{e7}) and (\ref{e1}) that $n_d^* = 
512/27$
 and
\begin{equation}
\Phi _d \simeq \frac{{9\pi }}{{128}}n_i^* . \label{e107}
\end{equation}
Thus, Eq.~(\ref{e4}) may have two solutions that 
correspond to two branches defining the parameters of a 
dust cloud. The junction of these branches is similar to the 
critical point characterized by the critical ion, electron, and 
dust particle density $n_{ic}^* $, $n_{ec}^* $, and 
$n_{dc}^* $, respectively. It is natural to associate this 
critical point with the void--dust boundary.

\begin{figure}
\includegraphics[width=9cm]{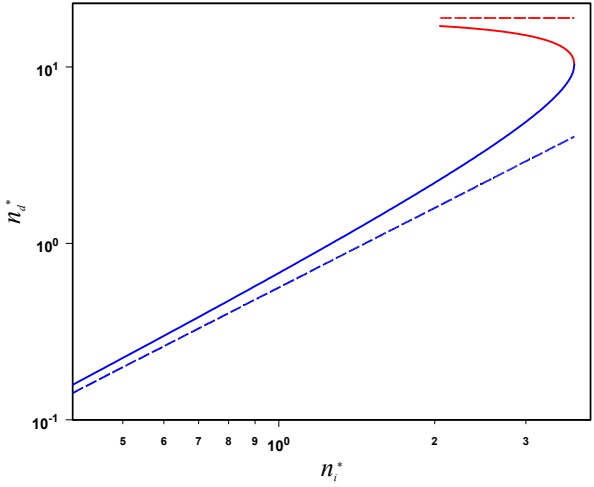}
\vskip-3mm
\caption{\label{f1}(Color online) Dust particle number 
density as a function of the ion number density. Dashed 
lines indicate approximate solutions for the low-density 
(\ref{e005}) and the high-density branch $n_d^* = 512/27$
}
\end{figure}

\begin{figure}
\includegraphics[width=9cm]{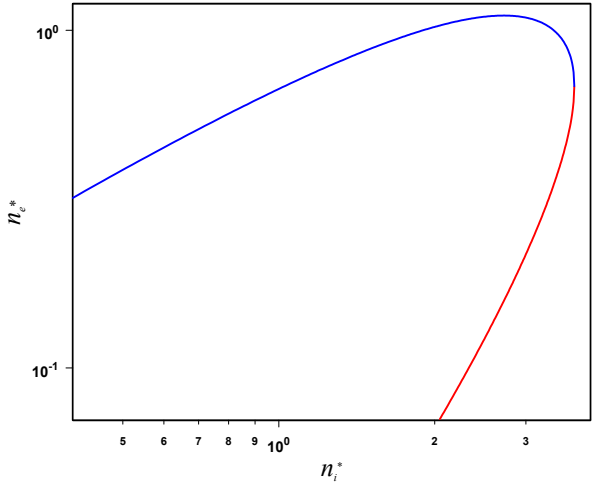}
\vskip-3mm
\caption{\label{f2}(Color online) Electron number density 
as a function of the ion number density}
\end{figure}

\begin{figure}
\includegraphics[width=9cm]{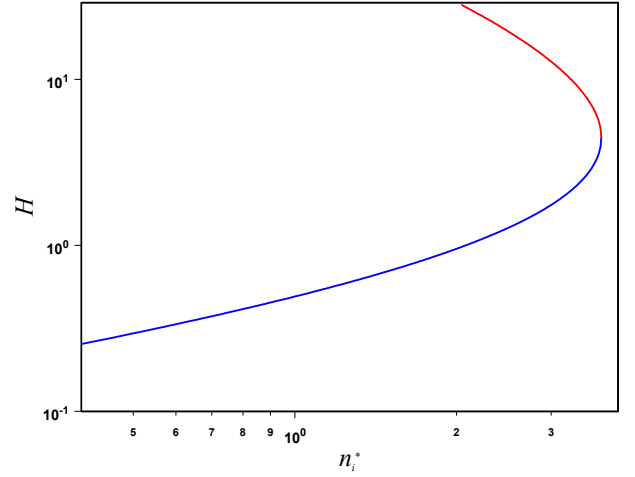}
\vskip-3mm
\caption{\label{f3}(Color online) Havnes parameter as a 
function of the ion number density}
\end{figure}

\begin{figure}
\includegraphics[width=9cm]{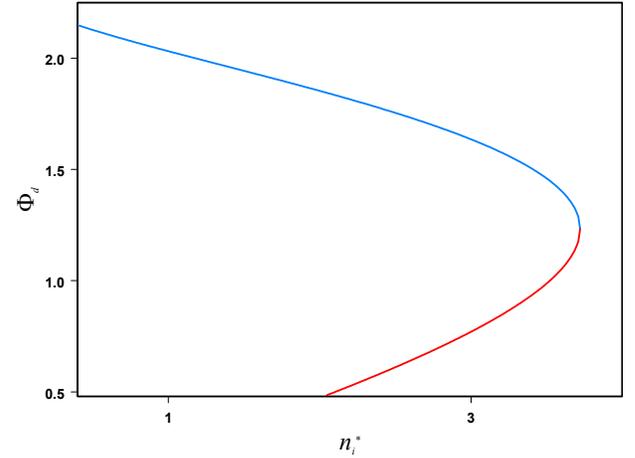}
\vskip-3mm
\caption{\label{f4}(Color online) Potential of a dust 
particle as a function of the ion number density}
\end{figure}

Figures~\ref{f1}--\ref{f4} show solutions of Eq.~(\ref{e4}) 
for the typical parameters $\tau = 135$
 and $\mu = 7.28 \times 10^4 $. Each dependence has two 
branches corresponding to a low (blue lines) and a high (red 
lines) dust number density. Both modes can in principle be 
observed depending on the RF discharge mode, but the 
low--density branch is positively most likely to be realized 
in experiment. It is seen in Fig.~\ref{f1} that the 
approximations of the lower branch (\ref{e005}) and of the 
upper branch $n_d^* = 512/27$
 are in a satisfactory correspondence with the exact 
solution. The branches of dependence $n_e^* (n_i^* )$
 are shown in Fig.~\ref{f2}. Note that at the upper branch in 
the vicinity of a critical point, this dependence is 
decreasing. This means that in this region, the electric field 
strength ${\bf{E}} = (T_e /e)(\nabla n_e /n_e )$
 changes its sign along with the direction of ion streaming. 
The possibility of such a peculiarity in the vicinity of the 
void--dust boundary was noted in Ref.~\onlinecite{38}. 
However, this does not affect the force balance equation 
(\ref{e1}) because it is independent of ${\bf{E}}$. The 
electron number density defines the Havnes parameter 
(Fig.~\ref{f3}). Along the lower branch sufficiently far 
from the critical point $H \ll 1$, therefore, the effect of a 
dust cloud on the electron and ion distributions can be 
neglected. At the critical point, the Havnes parameter 
reaches its maximum $H = 4.45$, so that the effect of dust 
particles on formation of the void--dust boundary must be 
significant.

Figure~\ref{f4} demonstrates that the dependence $\Phi _d 
(n_i^* )$
 is almost symmetric relative to the axis $\Phi _d = \Phi _c 
$, where $\Phi _c$ is the critical particle potential. 
Therefore, $\Phi _c \simeq \Phi _0 /2$. We substitute this 
into (\ref{e4}) with due regard for (\ref{e006}) to obtain 
the estimate
\begin{equation}
\begin{array}{*{20}c}
  {n_{ic}^* = \frac{{64}}{{9\pi }}\left[ {1 - 
\frac{1}{2}\left( {\frac{{\tau \Phi _0^2 }}{\mu }} 
\right)^{1/4} } \right]^2 \Phi _0 } \\
  { \approx \frac{{64}}{{9\pi }}\Phi _0 ,} \\
\end{array} \label{e007}
\end{equation}
since under typical experimental conditions, $(\tau \Phi 
_0^2 /\mu )^{1/4} /2 \ll 1$. We arrive at the conclusion that 
the critical ion number density $n_{ic} \approx (64/9\pi 
)(a_d T_e \Phi _0 /e^2 \lambda _{in}^3 )$
 is {\it proportional to the dust particle radius\/}.

An estimation for the critical electron density follows from 
(\ref{e6}) with $\Phi _d = \Phi _0 /2$
 and (\ref{e007}):
\begin{equation}
\begin{array}{*{20}c}
  {n_{ec}^* = \frac{{32}}{{9\pi }}\left( {\frac{{\tau \Phi 
_0^6 }}{\mu }} \right)^{1/4} \left[ {1 - \frac{1}{2}\left( 
{\frac{{\tau \Phi _0^2 }}{\mu }} \right)^{1/4} } \right]^2 } 
\\
  { \approx \frac{{32}}{{9\pi }}\left( {\frac{{\tau \Phi _0^6 
}}{\mu }} \right)^{1/4} .} \\
\end{array} \label{e008}
\end{equation}
As is seen in Fig.~\ref{f1}, the upper branch of the 
dependence $n_d^* (n_i^* )$
 is roughly approximated by a constant $512/27$, therefore 
$n_{dc}^* \sim 512/27$, whence it follows that the critical 
dust number density $n_{dc} \approx (128/9\pi )\lambda 
_{in}^{ - 3}$ is {\it independent of the dust particle 
radius\/}.

We expect $n_i$ to decrease with the increasing distance 
from the void center, so that the key assumption of this 
study, $a_d \tau \Phi _d \ll r_d $, flaws at some sufficiently 
large distance. The limit of validity for the proposed model 
is defined by the condition $a_d \tau \Phi _p = (3/4\pi n_d 
)^{1/3} $. For the case $H \ll 1$, we use (\ref{e005}) and 
$\Phi _p \simeq \Phi _0$ to find the lower bound for $n_i^* 
$,
\begin{equation}
n_i^* = \frac{2}{\pi }\left( {\frac{{\lambda _{in} }}{{\tau 
a_d }}} \right)^2 \Phi _0^{ - 1} . \label{e009}
\end{equation}

Another limitation comes from the Debye screening not 
taken into account in the foregoing. It was 
demonstrated\cite{49} that the approximation of the 
unscreened Coulomb potential results in a correct rate of 
the momentum transfer between the streaming ions and an 
{\it isolated\/} particle if the ion closest approach to the 
particle is set to be equal to the ion Debye length. This 
leads to the maximum impact parameter $\rho _{\max } = 
\lambda _{D_i } (1 + 2\Phi _d \tau a/\lambda _{D_i } 
)^{1/2} > \lambda _{D_i } $. For the Wigner--Seitz cell, 
the maximum impact parameter is of the order of $r_d $. 
Therefore, the Debye screening can be neglected if $\rho 
_{\max } > a_d \tau \Phi _p $. For typical experimental 
conditions, $a \sim 10^{ - 4} \;{\mbox{cm}}$
 and the Debye length corresponding to the minimum ion 
number density is $\lambda _{D_i } \sim 10^{ - 2} 
\;{\mbox{cm}}$
 (Sec.~\ref{s4}). Thus, $2\Phi _0 \tau a/\lambda _{D_i } > 
1$
 and $\rho _{\max } \simeq (2\Phi _d \tau a\lambda _{D_i } 
)^{1/2} $. With $n_i^*$ from (\ref{e009}), we have
\begin{equation}
\frac{{\rho _{\max } }}{{a_d \tau \Phi _0 }} \simeq \left( 
{\frac{{\lambda _{in} }}{{\lambda _{D_i } }}} 
\right)^{1/2} . \label{e091}
\end{equation}
For $\tau$ and $\mu$ treated above and $\lambda _{in} = 
2.1 \times 10^{ - 2} \;{\mbox{cm}}$
 under experimental conditions, the rhs of Eq.~(\ref{e091}) 
is greater than unity, i.e., $\rho _{\max } > r_d $. This 
means that the Debye screening can be neglected in this 
case. However for higher carrier gas pressures, the ratio 
$\lambda _{in} /\lambda _{D_i }$ could be less than unity, 
so that the lower bound for $n_i^*$ was defined by the 
Debye screening rather than by the particle potential 
overlapping. At the same time since $n_{ic}^*$ 
(\ref{e007}) is always greater than the minimum ion 
number density (\ref{e009}), the effect of the Debye 
screening in the vicinity of a critical point can be neglected 
in most cases. The range of parameters where the particle 
potential overlapping dominates is extended if we take into 
account that in the nonequilibrium gas discharge plasma, 
the length of Debye screening is larger than $\lambda 
_{D_i } $.\cite{51}

For $H \gg 1$, Eq.~(\ref{e107}) and the approximation 
$n_d^* \simeq 512/27$
 yield a lower bound for the high-density branch
\begin{equation}
n_i^* = \frac{{16}}{{3\pi }}\frac{{\lambda _{in} }}{{\tau 
a_d }}. \label{e109}
\end{equation}
The range of parameters in Figs.~\ref{f1}--\ref{f4} 
corresponds to the limitations (\ref{e009}) and 
(\ref{e109}). Since the high-density branch as a whole is 
not far from the critical point, the Debye screening can be 
neglected for it as well.

\section{\label{s3} THE RADIUS OF A CAVITY 
AROUND THE PROJECTILE}

Consider a large particle (projectile) with the radius $a_p$ 
in a dust cloud of smaller particles. Formation of a cavity 
around the projectile is quite similar to the cavity formation 
around a particle, which interacts with the molecules of 
surrounding liquid via a repulsive potential, known as 
self-trapping [see, e.g., Ref.~\onlinecite{42}]. Assuming 
that the electric filed of a charged projectile is screened 
outside the cavity, we can write the work of formation of a 
cavity with the radius $R_p$ as follows:
\begin{equation}
\begin{array}{*{20}c}
  {W(R_p ) = \frac{{4\pi }}{3}p_{\mathrm{st}} (R_p^3 - 
a_p^3 )} \\
  { + \frac{{Z_p^2 e^2 }}{2}\left( {\frac{1}{{R_p }} - 
\frac{1}{{a_p }}} \right),} \\
\end{array} \label{e8}
\end{equation}
where $p_{\mathrm{st}}$ is the static pressure of the 
particles in a dust cloud, $Z_p = (a_p T_e /e^2 )\Phi _p$ is 
the projectile charge, and $\Phi _p$ is its dimensionless 
potential. Note that the latter relation implies the neglect of 
screening of the projectile charge, which is justified in most 
cases. The first term on the rhs of (\ref{e8}) corresponds to 
the work against a constant pressure and the second one, to 
the energy of the electrostatic field induced by the projectile 
charge. In Eq.~(\ref{e8}), we neglect the surface tension 
term. The equilibrium cavity radius is defined by the 
condition $dW/dR_p = 0$, whence it follows that
\begin{equation}
R_p = \left( {\frac{{Z_p^2 e^2 }}{{8\pi p_{\mathrm{st}} 
}}} \right)^{1/4} . \label{e9}
\end{equation}
Relations like (\ref{e9}) are encountered, e.g., in the 
theories of positronium self-trapping in liquids.\cite{43} To 
find $p_{\mathrm{st}}$ we consider the limiting case $a_p 
\to a_d $. Obviously here $R_p \to r_d $, and the field of a 
particle is screened inside the Wigner--Seitz cell due to 
quasineutrality of the latter no matter how small (as 
compared to the cell radius) the screening length is. Thus, 
the model equation (\ref{e9}) is still valid in this case, and 
we readily derive from (\ref{e1}) and (\ref{e9})
\begin{equation}
p_{\mathrm{st}} = \frac{\pi }{{32}}(en_i \lambda _{in} 
)^2 . \label{e10}
\end{equation}

It is worth mentioning that, as it follows from (\ref{e9}) 
and the fact that $Z_p \propto a_p $, the relation $R_p 
\propto a_p^{1/2}$ can be treated as a {\it scaling law\/} 
for a projectile. If the cavity as a whole was neutral (as in a 
homogeneous system) the scaling would be different: $R_p 
\propto a_p^{1/3} $. Thus for a projectile, we note a 
violation of the overall quasineutrality of the cell, whose 
radius is defined by the pressure balance between the cell 
and surrounding dust cloud. The resulting positive excess 
charge may break the spherical symmetry of the cell due to 
the effect of the electric field. This may stipulate the 
projectile motion. The case of a projectile with a very large 
radius $a_p \gg \lambda _{in}$ is treated in Appendix. The 
only difference from the opposite case $a_p \ll \lambda 
_{in}$ treated above is the equation defining the projectile 
potential (\ref{e14}). The similarity between (\ref{e2}) and 
(\ref{e14}) is indicative of the fact that both cases are 
qualitatively similar.

\section{\label{s4} THE ANALYSIS OF AVAILABLE 
EXPERIMENTS}

We start the analysis with the data on the dust clouds 
formed by particles with the same diameter. It is a special 
problem, which branch of the solution of Eq.~(\ref{e4}) 
should be associated with experiment. We suppose that the 
region far from the critical point along the branch with a 
high particle number density can hardly be realized under 
any RF discharge mode because in this case, the rate of the 
ion--electron recombination on the particle surface would 
be too high and, due to the low electron number density, the 
ionization rate would be too low to sustain the discharge. 
However, this branch can be realized in the vicinity of the 
critical point in the unstable regime known as the heartbeat 
oscillations.\cite{44} In any case, the analysis of stability of 
the high-density branch is a separate problem to be 
addressed in future. Thus, we confine ourselves with the 
low-density branch and assume that far apart from the 
critical point, the effect of dust particles on the ion number 
density can be neglected.

Since the dust particle number density is usually estimated 
in the middle of a dust cloud, i.e., far apart from the 
void--dust boundary or from the critical point, it is 
reasonable to set approximately $\Phi _d \simeq \Phi _0 $. 
Since $\lambda _{in} \propto 1/p$, where $p$
 is the pressure of neutral gas atoms, and one can roughly 
assume that $n_i \propto p$,\cite{48} the product $n_i 
\lambda _{in}$ is almost constant. Then it follows from 
(\ref{e1}) that the ratio
\begin{equation}
\kappa = \frac{{r_d^2 }}{{a_d T_e }} \label{e20}
\end{equation}
must assume close values in different experiments. Thus, 
(\ref{e20}) is a scaling law for different dust clouds in the 
same carrier gas. Four sets of data of experiments 
performed in a wide range of argon pressures and particle 
diameters support this conclusion (Table~\ref{t1}): $\kappa 
= 0.209 \pm 0.04\;{\mbox{cm/eV}}$.
\begin{table*}
\caption{\label{t1} Parameters of the complex plasma in 
argon RF discharge and the corresponding ``dust invariant" 
$\kappa = r_d^2 /a_d T_e $.}
\begin{ruledtabular}
\begin{tabular}{cccccc}
$p,\;{\mbox{Pa}}$
 & $2a_d ,\;10^{ - 4} {\mbox{cm}}$
 & $n_d^{ - 1/3} ,\;10^{ - 4} {\mbox{cm}}$
 & $T_e ,\;{\mbox{eV}}$
 & $\kappa ,\;{\mbox{cm/eV}}$
 & Reference \\
\hline
16 & 1.55 & 114 & 3.8 & 0.170 & \onlinecite{28} \\
16 & 2.55 & 152 & 3.8 & 0.184 & \onlinecite{28} \\
30 & 9.55 & 370 & 4.5 & 0.245 & \onlinecite{11} \\
10 & 2.55 & 165 & 3.5 & 0.235 & \onlinecite{19} \\
\end{tabular}
\end{ruledtabular}
\end{table*}

Since $Z_d \sim a_d $, we can deduce from (\ref{e20}) that 
the maximum impact parameter for the ion--particle 
collision, which in our model is of the same order as $r_d $, 
must be proportional to $Z_d^{1/2}$ rather than to $Z_d $. 
The same dependence was obtained in 
Ref.~\onlinecite{49}, where the effect of the Debye 
screening on the ion--particle scattering cross section was 
investigated for an isolated particle.

We can estimate the ratio $u_i /v_{T_i }$ for experimental 
conditions listed in Table~\ref{t1}. If we estimate the ion 
flux velocity as $u_i = eE\lambda _{in} /m_i v_{T_i } $, 
where the electric field strength $E \simeq T_e /eL$
 and $L$
 is the discharge length scale, then $u_i /v_{T_i } \simeq 
\tau \lambda _{in} /L$. Adopting a typical value $L \sim 
1.5\;{\mbox{cm}}$
 we conclude that the sought ratio is confined in the interval 
$0.6 < u_i /v_{T_i } < 1.8$. This validates our basic 
relation (\ref{e1}).

Consider the experiments with argon as a carrier 
gas\cite{19} performed in the PK-3 Plus Laboratory 
onboard the ISS under microgravity conditions. The details 
on the setup can be found in Ref.~\onlinecite{18}. Dust 
particles injected into the main plasma with dispensers 
formed a cloud around the center of the chamber. A laser 
beam expanded to a light sheet was used for visualization 
of particle positions. Glow of particles illuminated by the 
laser sheet was recorded using high-resolution cameras, 
which make it possible to gain the most detailed 
information concerning a dust cloud. For the gas pressure 
and temperature of 10 Pa and $T_n = 300\;{\mbox{K}}$, 
respectively, we have $m_i = 6.63 \times 10^{ - 23} 
\;{\mbox{g}}$
 and $v_{T_i } = 2.5 \times 10^4 \;{\mbox{cm/s}}$. A dust 
cloud was formed by the melamine-formaldehyde particles 
with the radius $a_d = 1.275 \times 10^{ - 4} 
\;{\mbox{cm}}$
 and the mass $M_d = 1.31 \times 10^{ - 11} 
\;{\mbox{g}}$.\cite{19,22}

Some larger particles present in the chamber as well get 
sporadically accelerated and penetrate into the cloud, thus 
forming projectiles.\cite{26} According to 
Ref.~\onlinecite{22} projectiles are large particles with the 
radius $a_p = 7.5 \times 10^{ - 4} \;{\mbox{cm}}$. We 
can propose the following mechanism of projectile 
formation. Large particles not evacuated from the chamber 
after previous experiments are collected in agglomerates 
containing several particles. Upon illumination by the laser 
sheet, agglomerated particles are heated and deformed 
(estimates show that a typical thermal expansion 
deformations can be of the order of tens of the 
intermolecular spacing in the particle substance). This 
deformation is enough to entail cracking of the 
agglomerates. Then a detached particle is accelerated due to 
the Coulomb repulsion of the like charges $Z_p e$. For a 
pair of agglomerated particles, the projectile velocity upon 
penetration into the cloud $u_p$ can be estimated from the 
relation $M_p u_p^2 /2 = Z_p^2 e^2 /2a_p $, where $M_p$ 
is the projectile mass. With $Z_p = (a_p T_e /e^2 )\Phi _p 
$, where $\Phi _p \approx \Phi _0 $, we arrive at
\begin{equation}
u_p = \frac{{T_e \Phi _0 }}{e}\left( {\frac{{a_p }}{{M_p 
}}} \right)^{1/2} . \label{e015}
\end{equation}
For treated experimental conditions,\cite{19} 
Eq.~(\ref{e015}) yields $u_p \approx 14\;{\mbox{cm/s}}$, 
which is close to the initial projectile velocity upon 
penetration into the cloud observed in this experiment. This 
justifies the proposed mechanism of a projectile 
acceleration. The collective motion of dust particles 
induced by the projectile travel through the cloud was 
studied in Refs.~\onlinecite{22,25,45}.
\begin{figure}
\includegraphics[width=7.5cm]{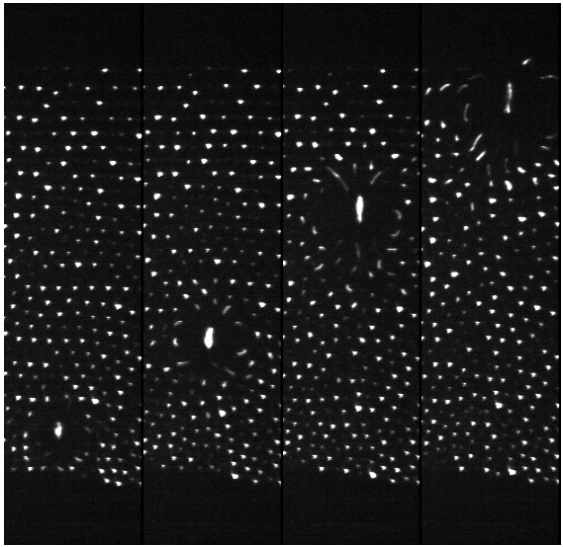}
\vskip-3mm
\caption{\label{f5} Snapshots (from left to right) of a 
projectile moving through the dust cloud in the upward 
direction toward the electrode, Ref.~\onlinecite{19}}
\end{figure}

\begin{figure}
\includegraphics[width=9cm]{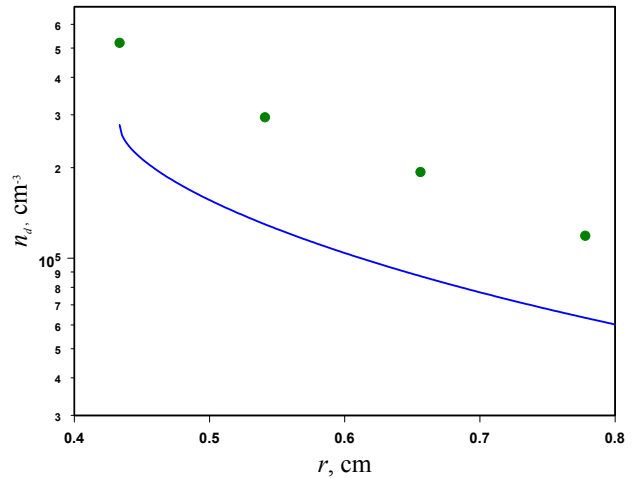}
\vskip-3mm
\caption{\label{f6}(Color online) Dust particle number 
density as a function of the distance from the center of a 
discharge. Solid line indicates the calculation using 
Eqs.~(\ref{e4}), (\ref{e5}), and (\ref{e070}); dots present 
processing of the snapshots\cite{19}}
\end{figure}

Successive frames of the dust cloud with a projectile 
moving slowly within it are shown in Fig.~\ref{f5}. A 
high-resolution camera allows one to resolve the change in 
the number density of dust particles. As is seen in 
Fig.~\ref{f5}, $n_d$ decreases with the increase of the 
distance from the void center. If we associate $r_0$ in 
(\ref{e070}) with the void--dust boundary (a critical point) 
then $n_{i0} = n_{ic} $, and we can use (\ref{e4}), 
(\ref{e5}), and (\ref{e070}) to calculate the spatial 
distribution of the particle number density. Calculation 
result for the particle number density distribution is 
compared with the data determined from the experiment in 
Fig.~\ref{f6}. In so doing, the experimental frames 
presented in Fig.~\ref{f5} were used for manual 
determination of $n_d = \Delta ^{ - 3} $, where $\Delta$ is 
the average visible interparticle spacing. Although this 
method of the number density determination is used by 
many authors, an appreciable error may be involved in it. 
The main sources of this error are unknown type of the 
crystal lattice and the angles between the vectors of a 
crystal lattice and the laser sheet plane. Apparently, this 
method of determination overestimates somewhat $n_d $. 
Nevertheless, Fig.~\ref{f6} is indicative of a satisfactory 
agreement between the theory and experiment.

The distribution of the complex plasma parameters are also 
illustrated by Figs.~\ref{f1}--\ref{f4}, which correspond to 
the parameters of the treated experiment. Note that the 
range of experimental parameters satisfies the condition 
(\ref{e009}), $n_d^* > 0.4$, defining the range of validity 
for the proposed theory. Also, it is worth mentioning that a 
rough estimate of the particle number density at the 
void--dust boundary $n_{dc} = (128/9\pi )\lambda _{in}^{ 
- 3} \approx 4 \times 10^5 \;{\mbox{cm}}^{ - 3}$ is close 
to the experimental value $5 \times 10^5 \;{\mbox{cm}}^{ 
- 3} $.\cite{19}

\begin{figure}
\includegraphics[width=9cm]{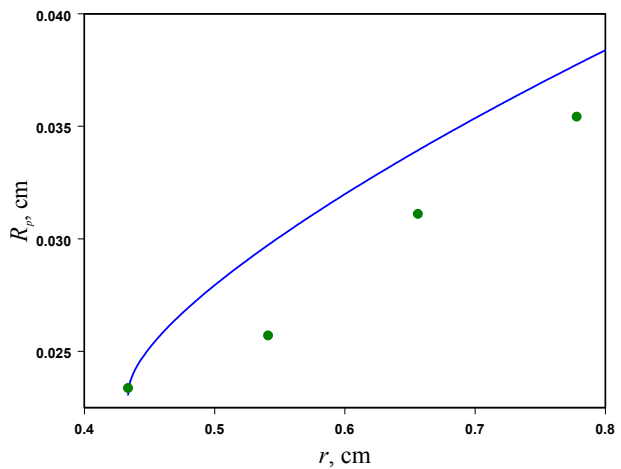}
\vskip-3mm
\caption{\label{f7}(Color online) Radius of a cavity around 
the projectile as a function of the distance from the center 
of a discharge. Solid line indicates the theoretical 
calculation using Eqs.~(\ref{e4}), (\ref{e070}), (\ref{e9}), 
and (\ref{e10}); dots denote processing of the 
snapshots\cite{19}}
\end{figure}

It can be seen in Fig.~\ref{f5} that the radius of a cavity 
around the projectile increases with the increase of the 
distance from the discharge center. This dependence can be 
determined experimentally by measurement of the cavity 
radius in the snapshots shown in Fig.~\ref{f5}. Calculation 
of $R_p$ can be performed using Eqs.~(\ref{e4}), 
(\ref{e070}), (\ref{e9}), and (\ref{e10}). Figure~\ref{f7} 
illustrates a good correspondence between the theory and 
experiment.\cite{19} According to the estimate (\ref{e10}) 
the static pressure of dust particles in the middle of a dust 
cloud is $p_{\mathrm{st}} = 6.3 \times 10^{ - 7} 
\;{\mbox{Pa}}$, which correlates with the experimental 
determination of this quantity at the same point based on 
the threshold of the projectile cavity deformation 
($p_{\mathrm{st}} = 3.0 \times 10^{ - 7} \;{\mbox{Pa}}$
\cite{45}). The obtained value of $p_{\mathrm{st}}$ is in 
agreement with a theoretical estimate\cite{45} 
$p_{\mathrm{st}} = Z_p^2 e^2 n_d /2r_d \approx 5.4 
\times 10^{ - 7} \;{\mbox{Pa}}$
 as well.

If two types of particles with the radii $a_1$ and $a_2$ are 
injected into the RF discharge then a binary system is 
formed. We denote the corresponding particle number 
densities by $n_1$ and $n_2 $. Such a system reveals a 
tendency toward phase separation.\cite{47} At the 
boundary between two phases, the ion and electron number 
densities must be continuous. Hence, in the vicinity of a 
boundary, the potentials of both types of particles must be 
equal, and we obtain from (\ref{e1}) the scaling law for a 
binary system
\begin{equation}
\left( {\frac{{a_1 }}{{a_2 }}} \right)^{1/2} = \left( 
{\frac{{n_2 }}{{n_1 }}} \right)^{1/3} . \label{e15}
\end{equation}
Unfortunately, the accuracy of the available experimental 
data is insufficient for a direct comparison with the theory.

\section{\label{s5} CONCLUSION}

We propose an equation of state for a dense dust cloud in 
the RF discharge, which expresses the potential of charged 
particles as a function of the ion number density given the 
electron temperature and the mass of a carrier gas atom. 
The key assumption is the overlap of potentials of the dust 
particles, which scatter streaming ions. As a result, the ion 
drag force proved to be dependent on the number density of 
particles. The latter gives rise to a two-branch solution 
corresponding to the normal and the increased particle 
number densities. For the branch with a normal density, the 
equation of state allows one to deduce a scaling law for dust 
clouds in the RF discharge, which relates the particle 
number density in a cloud to the particle radius and the 
electron temperature. As for the branch with the increased 
density, we expect that its manifestation could be the 
heartbeat oscillations, while a non-stationary dust cloud is 
involved in a self-sustained oscillation from one branch to 
another one. The investigation of stability for both branches 
will be addressed in future.

A notion of self-trapping of charged particles in condensed 
matter was used to calculate the radius of a cavity around a 
large particle (projectile), which penetrates into the dust 
cloud. The static pressure of dust particles involved in the 
resulting expression was obtained from the equation of state 
for a dust cloud. With respect to this problem, the 
calculation of refined distributions of charged particles 
within the cell around a projectile would give a clue to 
understanding the projectile motion through the dust cloud.

A good correspondence with the available experimental 
data and theoretical estimations of other studies points to 
the relevance of the proposed approach. Despite this fact, 
some assumptions involved in the proposed theory are 
worth discussion. First, the lhs of Eq.~(\ref{e1}) utilizes a 
simplified estimate for the cross section of the momentum 
transfer from ions to a particle. However, due to the 
potential overlapping in a dense particle system, refinement 
of this cross section cannot result in a significant correction 
to the results. The correction that is more noticeable could 
result from taking into account a nonlinear dependence of 
the drag force on the ion stream velocity in the region $u_i 
\sim v_{T_i } $.

The main objective of the proposed theory is to account for 
properties of a dust cloud far apart from the void--dust 
boundary. Nevertheless, it is of interest to extend the theory 
toward the critical point. In this respect, interpretation of a 
decreasing dependence of the electron number density on 
the ion number density in a close vicinity of the critical 
point (Fig.~\ref{f2}) is an urgent problem. The 
approximation of the spatial ion density by the dependence 
(\ref{e070}) in the vicinity of a critical point is knowingly 
too crude because it does not take into account the 
difference between $n_i$ and $n_e$ at $H > 1$. A rigorous 
approach would imply solution of the Poisson equation 
instead of the plasma quasineutrality equation (\ref{e3}); 
the continuity equation (\ref{e050}) should be completed 
by the ionization and recombination terms. However, 
complexity of the elementary processes occurring in the gas 
phase and a deficient information makes a correct 
simulation of the complex plasma problematic. Another 
point that is worth a separate treatment is the probable 
enhancement of the ion flux to the particle surface due to 
the ion trapping near dust particles and the resulting 
increase in the frequency of collisions between ions and 
neutrals. This effect must substantially reduce the particle 
charge as compared to the OML approximation.\cite{28} 
We suppose that this problem needs further investigation 
and can only note that all calculation results discussed in 
the foregoing are hardly compatible with any 
approximation other than the OML.

\begin{acknowledgments}
This research is supported by the Russian Scientific 
Foundation Grant No.~14-12-01235.
\end{acknowledgments}

\appendix
\section{\label{sA} THE POTENTIAL OF A 
MACROSCOPIC PARTICLE IN THE PLASMA}

We use the model of a cavity around a large particle 
developed in Sec.~\ref{s3}. For a macroscopic particle, 
however, the OML approximation is invalid because $a_p 
\gg \lambda _{in} $, and we have to estimate the ion flux to 
the surface of a large particle using an opposite 
approximation, namely, the diffusion one. We will neglect 
screening of the particle charge inside the cavity, so that 
$Z_p = (a_p T_e /e^2 )\Phi _p $. In this approximation, the 
ion flux includes the diffusion and drift 
components,\cite{46}
\begin{equation}
j_ +  = - D\nabla n_ +  + u_ + n_ + , \label{e11}
\end{equation}
where $D \simeq v_{T_i } \lambda _{in}$ is the diffusion 
coefficient, $n_ +$ is the local ion number density inside 
the cavity, and $u_ +  \simeq (Z_p e^2 \lambda _{in} 
/v_{T_i } m_i r^2 )$
 is its local drift velocity at the distance $r$
 from the cavity center. We substitute (\ref{e11}) into the 
local stationary continuity equation 
${\mathop{\mbox{div}}\nolimits} j_ +  = 0$
 to derive
\begin{equation}
\frac{d}{{d\tilde r}}\left( {\tilde r^2 \frac{{dn_ + 
}}{{d\tilde r}} + n_ + } \right) = 0, \label{e12}
\end{equation}
where $\tilde r = r/a_p \tau \Phi _p $. The solution of 
Eq.~(\ref{e12}) with the obvious boundary conditions $n_ 
+ (a_p ) = 0$
 and $n_ + (\infty ) = n_i$ is
\begin{equation}
n_ + (\tilde r) = \frac{{n_i }}{{1 - e^{\tau \Phi _p } }}\left( 
{e^{\tilde r^{ - 1} } - e^{\tau \Phi _p } } \right). 
\label{e13}
\end{equation}
Therefore, the ion flux on the particle surface is $j_ + (a_p ) 
= v_{T_i } \lambda _{in} n_i \tau \Phi _p /a_p $. It proves 
to be not much different from that given by the OML 
approximation. Indeed, the collisions with neutrals slow 
down the ions but they remove the energy and momentum 
limitations on the ions involved in the OML. These two 
factors seem to attenuate each other. Assuming the 
Boltzmann distribution for the electrons, we write the 
electron flux as $j_ - (a_p ) = (2\pi )^{ - 1/2} n_e v_{T_e } 
\exp ( - Z_p e^2 /a_p T_e )$. In a stationary state, $j_ + 
(a_p ) = j_ - (a_p )$, therefore, $\Phi _p$ is defined by the 
equation
\begin{equation}
\Phi _p e^{\Phi _p } = \left( {\frac{\mu }{{2\pi \tau }}} 
\right)^{1/2} \frac{{a_p }}{{\lambda _{in} }}\frac{{n_e 
}}{{n_i }}. \label{e14}
\end{equation}

Combination of (\ref{e14}), (\ref{e9}), and (\ref{e10}) 
makes it possible to estimate the radius of a cavity around a 
macroscopic particle in a dust cloud.
\providecommand{\noopsort}[1]{}\providecommand{\singleletter}[1]{#1}%
\end{document}